\documentclass[preprint,3p,number]{elsarticle}
\usepackage{comment}
\usepackage{xcolor}
\usepackage{subfigure}
\usepackage{graphicx}
\usepackage{array}
\usepackage{paralist}
\PassOptionsToPackage{hyphens}{url}\usepackage{hyperref}
\usepackage{siunitx}

\title{Design of high-strength, radiopure, electroformed copper–based alloys for rare-event searches: Impact of layer configuration on heat treatments}
\author{D. Spathara$^a$}
\author{P. Knights$^a$}
\author{K. Nikolopoulos$^{a,b}$}
\ead{d.spathara@bham.ac.uk}
\affiliation{organization={School of Physics and Astronomy, University of Birmingham},
postcode={B15 2TT},
state={Birmingham},
country={United Kingdom}}
\affiliation{organization={Institute for Experimental Physics, University of Hamburg},
postcode={22761},
state={Hamburg},
country={Germany}}

\begin{document}

\begin{abstract}
State-of-the-art and next-generation rare-event search experiments rely on detector materials with stringent requirements on radiopurity and mechanical performance. Additive-free electroformed copper offers exceptional radiopurity, but is limited in mechanical strength, motivating the exploration of application-specific copper-based alloys. Early investigations, based on direct experimentation, explored the synthesis of CuCr alloys through electrodeposition and thermal processing. Subsequently, modeling tools based on the thermodynamic and kinetic properties of the alloy compositions were employed, which led to specific proposals for improved thermal processing. Moreover, the systematic application of computational thermodynamics to materials design further motivated the investigation of CuCrTi alloys, in addition to CuCr alloys. This materials design approach has shaped a trajectory towards designing high-performance, radiopure copper-based alloys, minimizing lengthy and costly trial-and-error. In this work, we explore the impact of initial layer configuration on the effectiveness of heat treatments, paving the way toward manufacturable, radiopure, multicomponent alloys for future low-background experiments.

\end{abstract}


\maketitle

\section{Introduction}
Searches for rare-event phenomena —such as the direct detection of dark matter and detection of neutrinoless double-$\beta$ decays— drive the need for experiments with extremely low backgrounds~\cite{CUORE:2017tlq,NEWS-G:2020fhm,XLZD:2024nsu}. At the heart of this effort lies the development of detector materials that combine exceptional radiopurity with high-end mechanical performance. Additive-free electroformed copper (EFCu) has been established as a material of choice, delivering ultra-low levels of radioactive contaminants and enabling groundbreaking advances in experiment sensitivity~\cite{hoppe2008use,Hoppe:2014nva,MAJORANA:2016lsk}. However, the mechanical limitations of pure EFCu, particularly its high ductility and modest yield strength, present significant constraints for its deployment in detectors requiring structural integrity under high loads or pressures~\cite{osti_1039850}.

This challenge has motivated an ongoing effort to develop application-specific copper-based alloys that combine the radiopurity of EFCu with enhanced mechanical properties. Initial experimental investigations demonstrated that introducing small quantities of chromium (Cr) to Cu, via additive-free electrodeposition followed by thermal processing, can result in substantial increases in mechanical strength while maintaining high radiopurity.
Specifically, Cr concentrations in the range of 0.3-0.58~wt\%, combined with heat treatment and aging, can significantly increase the hardness of EFCu by 70-100\%~\cite{Suriano:2018nrb,Vitale:2021xrm,osti_1039850}.  Moreover, in Ref.~\cite{Suriano:2018nrb} the projected radiopurity of the CuCr alloy with 0.585~wt\% is at levels comparable to those of EFCu. These advances, however, were achieved through direct experimentation, limiting the possibility for homogeneous alloy composition and process optimization.

Recent work has leveraged computational thermodynamics to address this limitation, opening a pathway to systematic alloy design for rare-event searches~\cite{Spathara:2025bfw}. Specifically, CALPHAD-type methods, kinetic simulations using DICTRA and precipitation modeling via TC-PRISMA, a predictive framework for the manufacture and processing of high-purity Cu–Cr alloys was established~\cite{Spathara:2025bfw}. By systematically modeling diffusion, phase transformations, and microstructure evolution during solution heat treatment and aging, these studies identified optimized thermal processing parameters and layer configurations that yield homogeneous, mechanically robust alloys within practical timescales. The model was validated through direct comparison with the available experimental measurements for Cr/Cu layer configuration. 

Building on these developments, the approach was systematically applied at a broader scope~\cite{Spathara:2025hrp}. Mutlicomponent systems were considered, notably CuCrTi alloys, inspired by the promise of further mechanical enhancement~\cite{HUANG2021102378}. Through phase diagram analysis and diffusion modeling, these investigations outlined the thermodynamic and practical challenges of incorporating titanium (Ti)—particularly the formation of intermediate phases and the need for controlled layer arrangements and heat-treatment conditions. Moreover, the broader impact of these advances in materials was examined, 
demonstrating that these can lead to enhanced sensitivity of future rare-event direct search experiments such as DarkSPHERE~\cite{NEWS-G:2023qwh,Knights:2025ogz}, and XLZD~\cite{XLZD:2024nsu}.

A key aspect shaping the manufacturability and effectiveness of Cu-based alloys is the role of initial layer in the sequential electrodeposition and subsequent thermal processing of the alloy. The spatial arrangement and relative thickness of these layers set the foundation for alloy homogenization, phase stability, and microstructural development during heat treatments. In this work, the impact of initial layer thickness and configuration on the effectiveness of solution heat treatment in the synthesis of high-strength, radiopure, electroformed Cu-based alloys is investigated. The parameter space linking layer arrangements to alloy homogenization is mapped, which enables the identification of manufacturable configurations that optimize microstructural control and mechanical enhancement.

\section{Results and Discussion}
The methods employed in this work are detail in Refs.~\cite{Spathara:2025bfw, Spathara:2025hrp}.
In the context of this work, two methods of thermal processing are considered: 
\begin{inparaenum}[a)]
\item solution heat treatment; and
\item aging.
\end{inparaenum}
The former aims at a homogenized alloy composition, while the latter aims at precipitation strengthening and follows the solution heat treatment. 
The modeling using computational thermodynamics aims to determine the optimal time and temperature for these thermal processing stages. These are developed within the CALPHAD framework (CALculation of PHAse Diagrams)~\cite{lukas2007computational, SPENCER20081, kattner2016calphad}. In this work, the CALPHAD-type databases Thermo-calc Software TCHEA6 and the MOBHEA3 High Entropy alloys database~\cite{TCdatabase1, TCdatabase6} are used for the description of the thermodynamic and kinetic properties of Cr-Cu and Cr-Cu-Ti systems. 

For the solution heat treatment stage, the DICTRA module is used~\cite{andersson2002thermo}, which solves the diffusion equations at the interface of two regions of different composition during phase transformation. This is performed at a single temperature for a specific length of time, i.e. an isothermal.

The nucleation and growth of precipitates during aging are simulated using the TC-PRISMA module~\cite{tang2018kinetic}, according to the Langer-Schwartz theory~\cite{langer1980kinetics}. In addition, the Kampmann Wagner numerical method is adopted~\cite{wagner2001homogeneous}. In this way, the nucleation and growth of the precipitating secondary phase are considered simultaneously. 

\subsection{Cr/Cu layer configurations and temperature dependence for forming CrCu alloys}
As discussed in detail in Ref.~\cite{Spathara:2025bfw}, maximum precipitation strengthening is achieved when the Cr concentration in the CuCr alloy reaches its maximum levels of solubility, which is approximately 0.58~wt\%. It was found that $1\,\si{\micro\meter}$  Cr in contact with $400\;\si{\micro\meter}$  Cu yielded a homogenized CuCr alloy with around 0.2~wt\% Cr. In contrast, a $10\;\si{\micro\meter}$ Cr layer, was found to be too thick to be fully incorporated in $400\;\si{\micro\meter}$ Cu in reasonable time duration. 

\begin{figure*}[h]
    \centering
    \subfigure[\label{fig:3a}]{\includegraphics[width=0.32\linewidth]{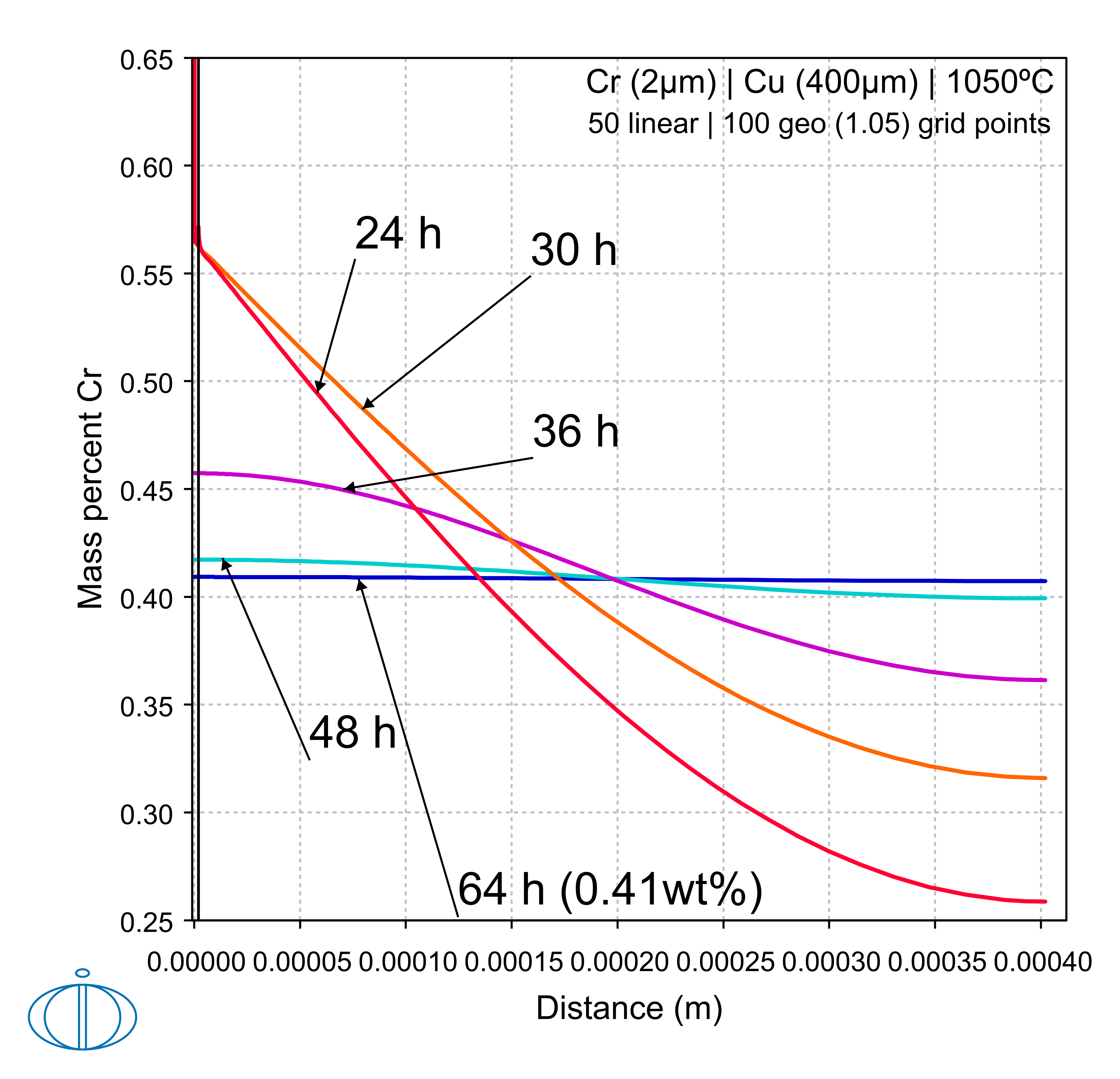}}
    \subfigure[\label{fig:3b}]{\includegraphics[width=0.32\linewidth]{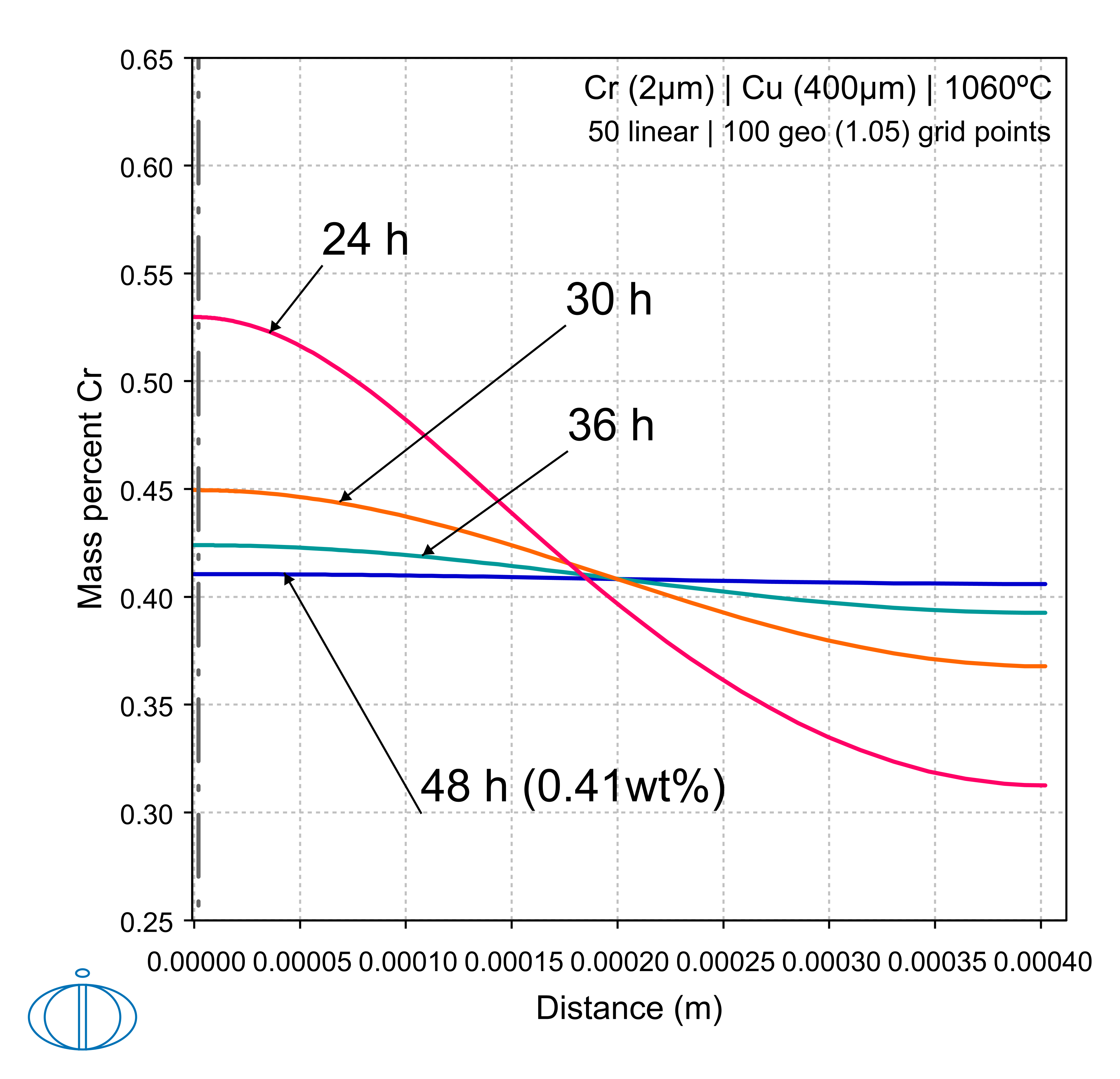}}
    \caption{Cr concentration profiles as a function of depth in the $400\;\si{\micro\meter}$ Cu region, in contact with $2\;\si{\micro\meter}$ Cr, at \subref{fig:3a} $1050\,^{\circ}$C and \subref{fig:3b} 1060$^{\circ}$C using 1D DICTRA simualtions.
    \label{fig:1}}
    \vspace{-0.3cm}
\end{figure*}

Solution heat treatment at two temperatures for $2\;\si{\micro\meter}$ Cr in contact with $400\;\si{\micro\meter}$ of Cu is modeled. Figure~\ref{fig:1} shows the Cr composition profile after simulation results using DICTRA at $1050\,^{\circ}$C and $1060\,^{\circ}$C. The results indicate that the increase in temperature by 10 degrees decreases substantially the time required for the Cu-0.41Cr alloy to be fully homogenized, from 64 to 48 hours at $1060\,^{\circ}$C, compared with results at temperature $1050\,^{\circ}$C.

 In Ref.~\cite{Spathara:2025hrp}, 3 options for the Cr/Cu layer configuration leading to the Cu-0.5Cr alloy composition were suggested:
 \begin{inparaenum}[a)]
     \item 1.5 in contact with $245\;\si{\micro\meter}$ Cu;
     \item 3 $\mu m$ Cr in contact with $245\;\si{\micro\meter}$ Cu, at each side; and
     \item 6 $\mu m$ Cr in contact with $490\;\si{\micro\meter}$ Cu, at each side.
 \end{inparaenum}
In the first 2 cases, 27 hours of solution heat treatment at $1050\,^{\circ}$C was the practical time predicted for alloy homogenization. Evidently, there are numerous cases of Cr/Cu layer configuration that would lead to homogenized CuCr alloys with the desired Cr content. Due to the limited solubility of Cr in Cu, the duration of solution heat treatment decreases with thinner Cr layer in thicker Cu layer.

\subsection{Cr/Ti layer configuration to obtain CrTi alloys}
In Ref.~\cite{Spathara:2025hrp}, a methodology to synthesize CuCrTi alloys was proposed, which would provide enhance mechanical thrength. In addition to the reported effect of Ti additions in alloys mechanical enhancement~\cite{HUANG2021102378, EZE2018163}, DICTRA simulations have suggested that Ti decreases the solution heat treatment time required for full alloy homogemization. The proposed method included the synthesis of CrTi alloy as the first stage of the manufacturing process. Subsequently, an EFCu layer would then be plated to this. 
In order for the solution heat treatment of the CrTi alloy in contact with EFCu to take place at $1050\,^{\circ}$C, the Ti content in the CrTi alloy should not exceed 6.5~wt\%. Above this, incipient melting~\cite{pang2012solution, hegde2010designing,andilab2020situ} of the Ti could compromise radiopurity and mechanical properties. 

\begin{figure*}[h]
    \centering
    \subfigure[\label{fig:4a}]{\includegraphics[width=0.32\linewidth]{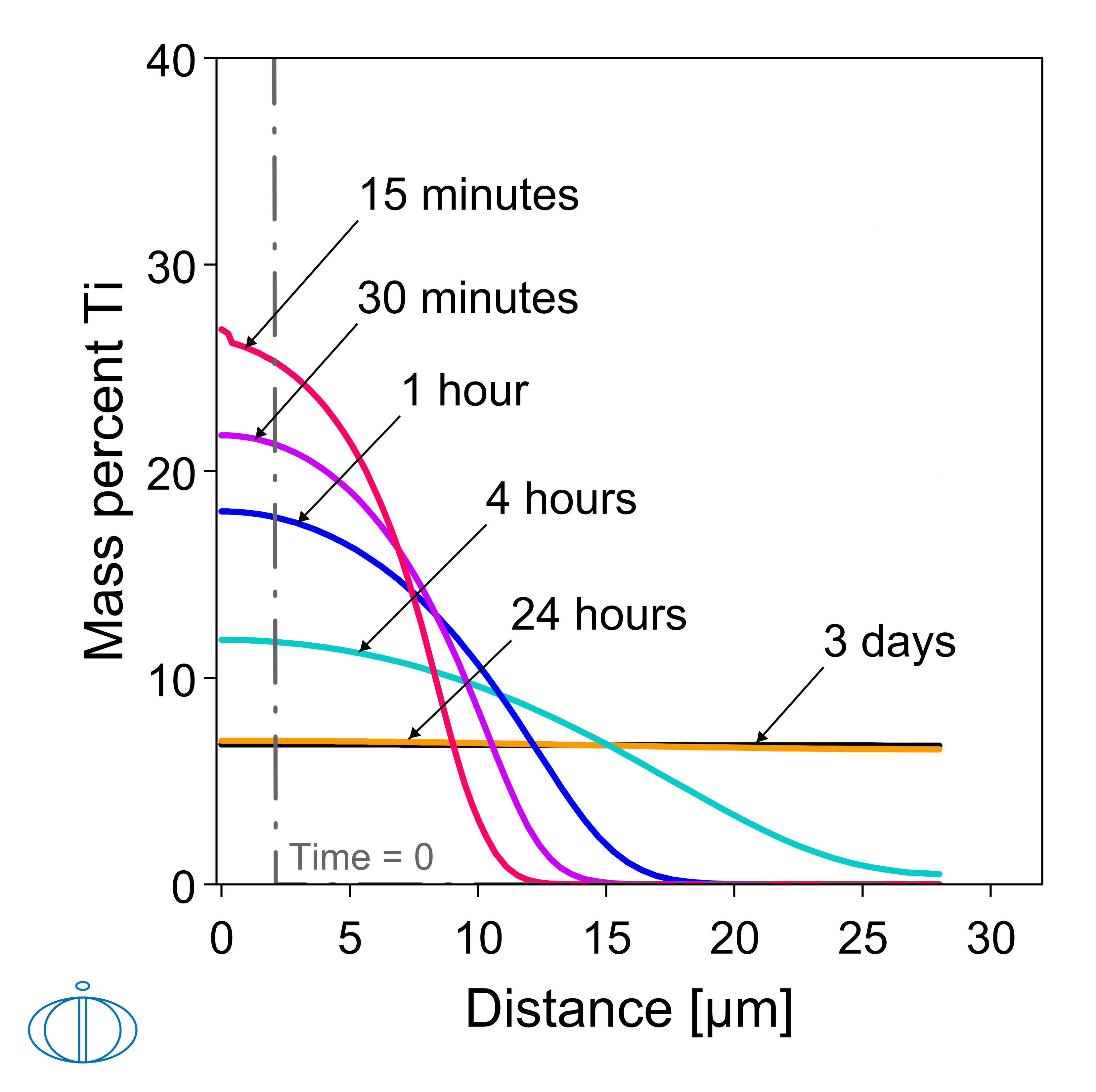}}
    \subfigure[\label{fig:4b}]{\includegraphics[width=0.32\linewidth]{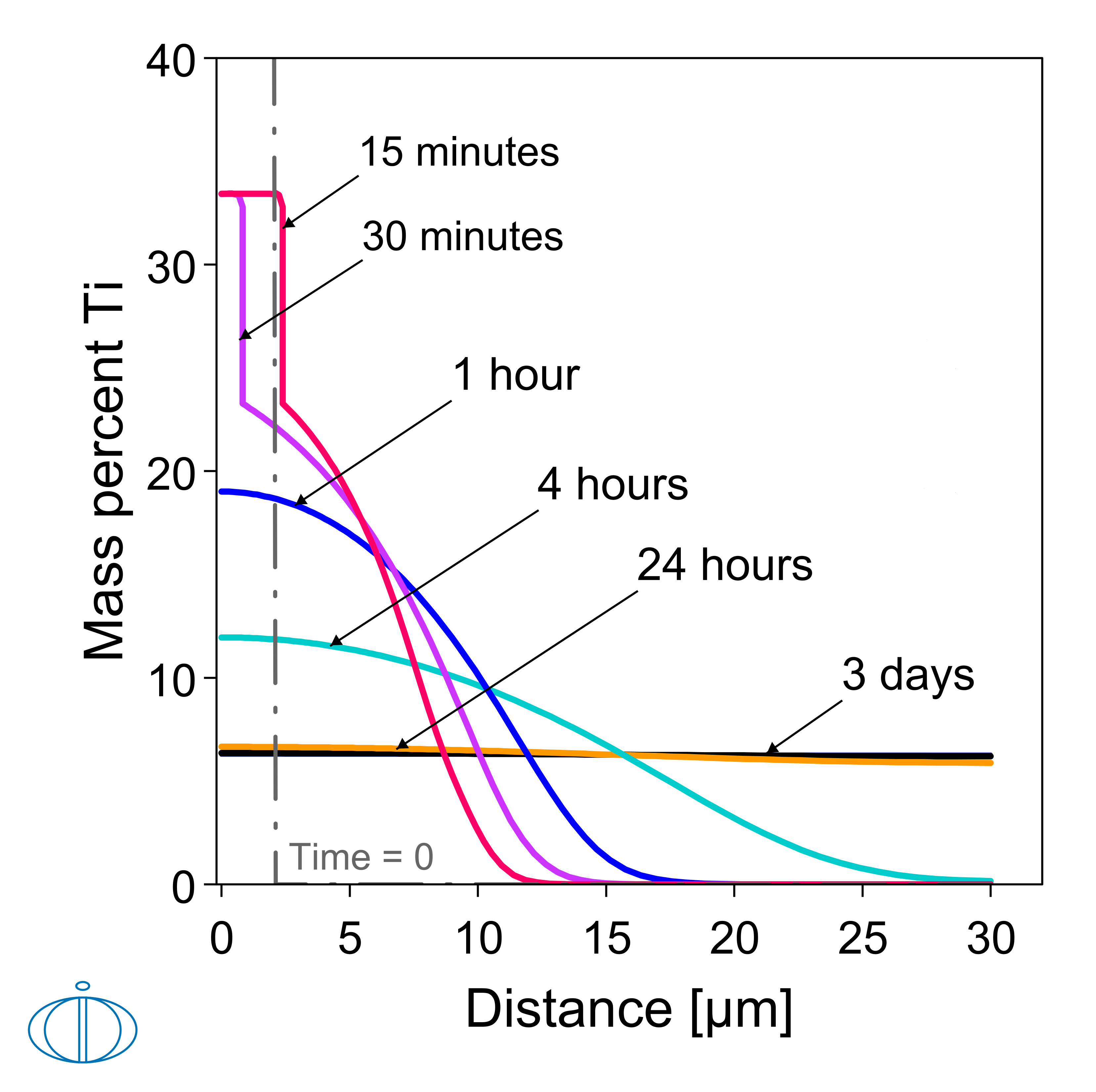}}
    \subfigure[\label{fig:4c}]{\includegraphics[width=0.32\linewidth]{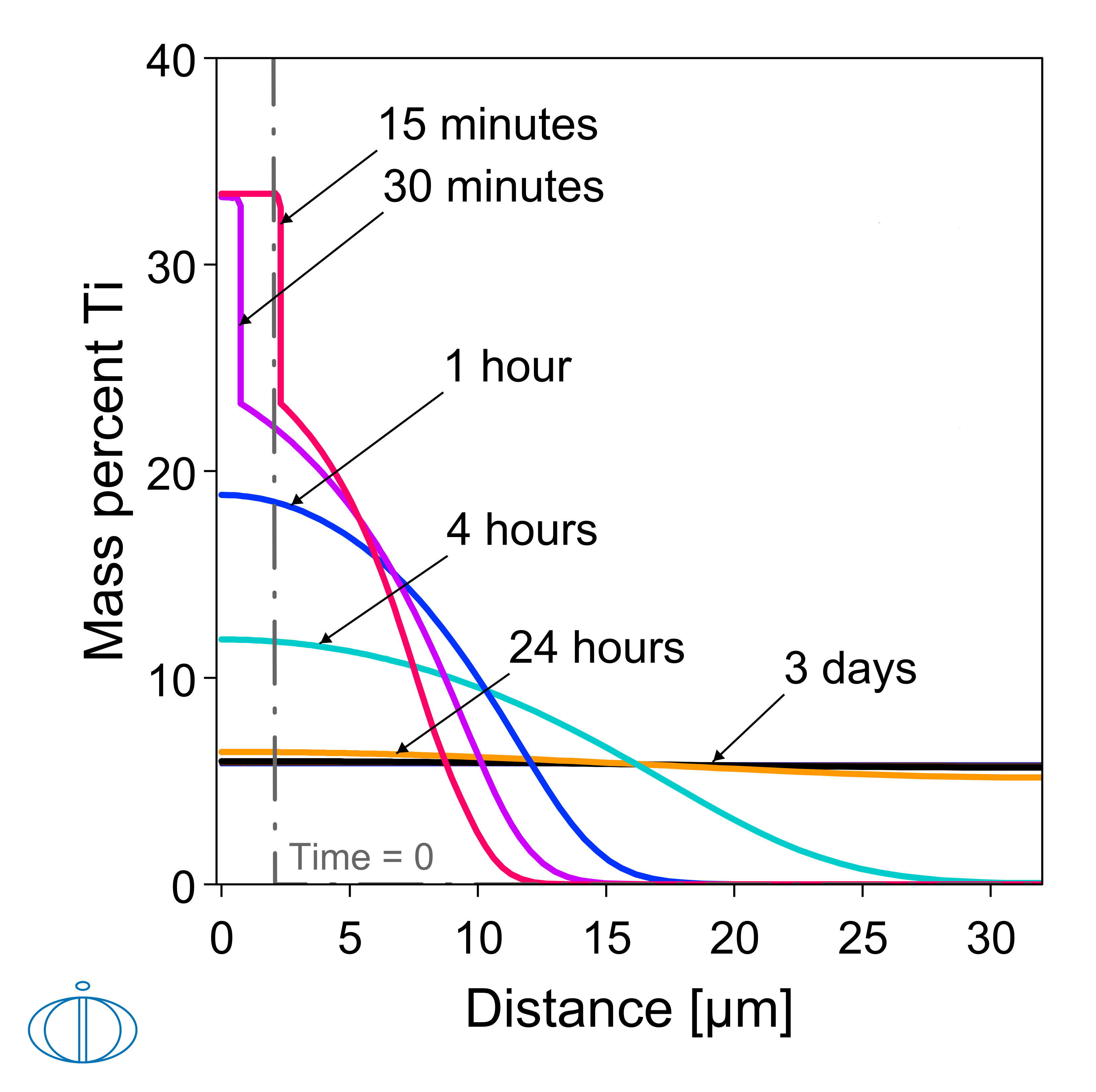}}
    \caption{Ti concentration profiles of $2\;\si{\micro\meter}$ Ti in contact with Cr of \subref{fig:4a} $26\;\si{\micro\meter}$, \subref{fig:4b} $28\;\si{\micro\meter}$ and \subref{fig:4c} $30\;\si{\micro\meter}$, leading to Cr-6.75Ti, Cr-6.28Ti and Cr-5.81Ti (in wt\%) at $1350\,^{\circ}$C, using 1D DICTRA simulations.
    \label{fig:4}}
    \vspace{-0.3cm}
\end{figure*}

Figure~\ref{fig:4} shows the results of DICTRA 1D simulations between $2\;\si{\micro\meter}$ Ti in contact with 26, 28 and 30$\;\si{\micro\meter}$ of Cr at $1350\,^{\circ}$C, resulting in a Ti content of 6.75~wt\% (Fig.\ref{fig:4a}), 6.28~wt\% (Fig.~\ref{fig:4b}) and 5.81~wt\% (\ref{fig:4c}). The former case would lead to a Ti concentration that exceeds 6.5~wt\%. In the other two cases, Ti content is below this threshold but sufficiently high for mechanical enhancement of the alloy.

\section{Conclusion}
Electroformed copper-chromium (CuCr) alloys, post-thermal processing, are stronger than EFCu and hold promise in terms of required radiopurity levels for rare-event searches experiments. Consequently, their development is crucial for next-generation rare-event detection experiments.

In addition, computational thermodynamics was used to study the thermal processing parameters of a range of different layer configurations for CuCr and CuCrTi alloys. This builds on the methodology proposed in earlier works which demonstrated the use of DICTRA and TC-PRIMA simulations to predict material properties and optimize the manufacturing process. It was demonstrated that the manufacturing of homogenised CuCr and CrTi alloys can be achieved in practical timescales.

While the Cr/Ti layer configuration has been studied in this work, a next step is to explore the CrTi/Cu layer configurations in terms of determining practical times for solution heat treatments that will lead to alloys with Cr content between 5-5.8~wt\%. 

Future work includes a systematic approach to explore the effect of the duration of solution heat treatments to the alloys homogeneity. A map of the relationship between time and thicknesses of Cr/Cu CrTi/Cu layer configurations aimed at specific alloy compositions will inform future experiments to validate the developed models.

Manufacturing and property characterization of the designed radiopure CuCr and CuCrTi alloys in this work will further determine the suitability of the commercial CALPHAD-type thermodynamic and kinetic databases used and will pave the way to the development of more accurate predictions and optimized thermal processing parameters for optimum mechanical enhancement.

\section{Declaration of competing interest}
The authors declare that they have no known competing interests or personal relationships that could appear to influence the reported work.

\section{Acknowledgement - Funding}
Valuable discussions with Dr. Andre Schneider from Thermo-Calc Software are acknowledged. 
Support from UKRI-STFC (No. ST/W000652/1) 
and the UKRI Horizon Europe Underwriting scheme (PureAlloys -- EP/X022773/1) is acknowledged.
The support of the Deutsche Forschungsgemeinschaft (DFG, German Research Foundation) under Germany’s Excellence Strategy -- EXC 2121 “Quantum Universe”-390833306 is acknowledged.

\bibliographystyle{elsarticle-num}
\bibliography{bibliography}
\end{document}